\documentclass[a4paper,10pt]{article}
\usepackage{amsmath,amssymb,amsthm,bbm,epsfig,graphics,graphicx,hyperref}
\usepackage{verbatim,braket,cleveref, enumerate, tensor}
\usepackage{color,subfigure}
\usepackage{authblk}
\newtheorem{theorem}{Theorem}
\newtheorem{lemma}[theorem]{Lemma}
\newtheorem{definition}[theorem]{Definition}

\newtheorem{example}[theorem]{Example}

\newcommand{\HH}{\ensuremath{\mathcal{H}} }
\newcommand{\PG}{\ensuremath{\mathcal{G}} }
\newcommand{\gG}{\ensuremath{\mathsf{G}} }
\newcommand{\cV}{\ensuremath{\mathcal{V}} }
\newcommand{\cE}{\ensuremath{\mathcal{E}} }
\newcommand{\ZZ}{\ensuremath{\mathbb{Z}} }
\newcommand{\CC}{\ensuremath{\mathbb{C}} }
\newcommand{\Ccal}{\ensuremath{\mathcal{C}} }
\newcommand{\CZ}{\ensuremath{\mathsf{CZ}}}
\newcommand{\AME}[1]{AME($#1$)}
\newcommand{\QSS}[2]{$((#1,#2))$ threshold QSS scheme}
\newcommand{\floor}[1]{\ensuremath{\lfloor #1 \rfloor}}

\newcommand{\Id}{\mathbbm{1}}
\newcommand{\vv}[1]{\ensuremath{\mathbf{#1}}}

\newcommand{\ketbraind}[3]{\ket{#1}\bra{#2}_{#3}}
\newcommand{\braketind}[3]{{\vphantom{\braket{#2}}}_{#1}\!\braket{#2}_{#3}}

\newcommand{\tth}{\ensuremath{\mathrm{{th}}}}

\DeclareMathOperator{\diag}{diag}
\DeclareMathOperator{\Tr}{Tr}

\crefname{equation}{Equation}{Equations}
\crefname{align}{Equation}{Equations}
\crefname{definition}{Defintion}{Defintions}
\crefname{item}{}{}

\title{Absolutely Maximally Entangled Qudit Graph States}
\author{Wolfram Helwig}
\affil{Center for Quantum Information and Quantum Control (CQIQC),\\
Department of Physics,
University of Toronto, Toronto, Ontario, M5S 1A7, Canada}

\begin{document}

\maketitle

\begin{abstract}
Absolutely maximally entangled (AME) states are multipartite entangled states that are maximally entangled for any possible bipartition. In this paper, we study the description of AME states within the graph state formalism. The graphical representation provides an intuitive framework to visualize the entanglement in graph states, which makes them a natural candidate to describe AME states.
We show two different methods of determining bipartite entanglement in graph states and use them to define various AME graph states. We further show that AME graph states exist for all number of parties, and that any AME graph states shared between an even number of parties can be used to describe quantum secret sharing schemes with a threshold or ramp access structure directly within the graph states formalism.
\end{abstract}

\section{Introduction}
Entanglement is one of the most valuable resources in quantum information processing. The more entangled states are, the more powerful are the operations that can be performed with them. However, for multipartite states, many different types of entanglement exist \cite{Bennett2000, Vidal2000, Dur2000}, and it is not always clear, which states are most suitable for a specific task. 

One possibility of quantifying multipartite entanglement is to look at the bipartite entanglement the state possesses for various partitions of the parties into two sets. States that maximize this entanglement criterion are absolutely maximally entangled (AME) states\cite{Helwig2012, Helwig2013}, which are states that are maximally entangled for every possible bipartition of the state. It was shown that these states exist for any number of parties if the system dimensions are chosen appropriately, and that they can be used for various quantum information tasks, like parallel teleportation of multiple states between arbitrary sets of parties, open-destination teleportation, and quantum secret sharing. Formally, there are different equivalent ways to ascertain that a state is absolutely maximally entangled. Which one to use often depends on the application under consideration.

\begin{definition}
\label{def:AME}
	An absolutely maximally entangled state is a pure state shared between $n$ parties $P=\{1,\ldots,n\}$, each having a system of dimension $d$, so a state $\ket{\Phi} \in \HH_1 \otimes \cdots \otimes \HH_n$ and $\HH_i \cong \CC^d$, with the following equivalent properties:
	\begin{enumerate}[(i)]
		\item $\ket{\Phi}$ is maximally entangled for any possible bipartition. This means that for any bipartition of $P$ into disjoint sets $A$ and $B$ with $A\cup B  = P$ and, without loss of generality, $m=|B|\leq |A|=n-m$, the state $\ket{\Phi}$ can be written in the form 
		\begin{equation}
			\label{eq:defAMEstate}
			\ket\Phi = 
			\frac{1}{\sqrt{d^m}}\sum_{k\in \ZZ_d^{m}} 
			\ket{k_1}_{B_1}\cdots \ket{k_{m}}_{B_{m}}
			\ket{\phi(k)}_A,
		\end{equation}
		with $\braket{\phi(k)|\phi(k')} = \delta_{kk'}$.
		\item The reduced density matrix of every subset of parties $A\subset P$ with $|A|\leq \frac{n}{2}$ is totally mixed.
		\label{def:AMEdensity}
		\item The von Neumann entropy of every subset of parties $A\subset P$ with $|A| \leq \frac{n}{2}$ is maximal, $S(A) = |A| \log d$.
\label{def:AMEentropy}
	\end{enumerate}
	We denote such a state as an \AME{n,d} state. Note that to check if an state is absolutely maximally entangled, it suffices to check the maximal entanglement for all bipartitions with $|A| = \floor{\frac{n}{2}}$, as the maximal entanglement for smaller bipartitions follows immediately.
\end{definition}

In this paper, we will use the graph state formalism to describe AME states. Graph states are a special class of stabilizer states, and have been introduced for qubits and qudits of prime dimension\cite{Briegel2001, Bahramgiri2006}. They offer a nice graphical representation of multipartite entangled states and have found its use in a variety of quantum information applications, like quantum computing \cite{Raussendorf2001}, error correction \cite{Schlingemann2002, Grassl2002, Looi2008, Beigi2011}, and quantum secret sharing \cite{Markham2008, Keet2010}.

We will show two methods for checking bipartite entanglement in graph states. One makes use of the intuitive graphical representation of graph states, while the other one allows to efficiently check if a graph state is absolutely maximally entangled, even for high dimensional systems and a large number of parties -- a task that is generally hard to accomplish in the Dirac notation, as it involves tracing over high-dimensional density matrices to verify the condition in Definition~\ref{def:AME}~(\ref{def:AMEentropy}).

Examples of AME graph states will be given, among others a previously unknown AME states for seven qutrits that we were able to find in computer searches that used the efficient method to determine bipartite entanglement in graph states. Further, we will show how the method presented in Ref.~\cite{Helwig2013} to construct graph states from classical codes can also be used to construct AME graph states for any number of parties. Given a certain graph state, it is straightforward to write down a quantum circuit, consisting of controlled-$Z$ gates that produces the graph state. Thus this method will enable us to write down a quantum circuit that creates an AME state for any number of parties, and once a method exists to experimentally implement controlled qudit gates, the approach in this paper provides a straightforward way to experimentally create qudit AME states. At this point, graph states have been experimentally created for up to six qubits \cite{Walther2005, Lu2007, Ceccarelli2009, Gao2010}.

Quantum secret sharing (QSS) with qudits has already been investigated before \cite{Markham2008, Keet2010}. However, only a few specific examples of graph states that can be used for QSS could be given, and the question which graph states are generally suitable for QSS has been left open. We answer this question by showing that all AME graph states shared between an even number of parties can be used to construct threshold QSS schemes \cite{Cleve1999}, as well as for QSS schemes with a more general \emph{ramp} access structure \cite{Blakley1984}. The connection between AME states and threshold QSS schemes has already been shown before \cite{Helwig2012, Helwig2013}, the treatment here is to show that the derivation of QSS schemes from AME states can also be completely described within the graph state formalism. The results of Ref.~\cite{Helwig2013} further show that AME graph states are the only graph states that result in threshold QSS schemes.

This paper is structured as follows. In Section~\ref{section:quditgraphs} we introduce qudit graph states and their representation as stabilizer states. In Section~\ref{section:entanglement} we show two different methods for checking the bipartite entanglement in graph states. Section~\ref{section:AMEstates} gives examples of AME states, which were found by using the methods presented in the previous section. We further show that AME graph states exist for any number of parties. In Section~\ref{section:QSS} we show how any AME state shared between even number of parties can be used to implement quantum secret sharing right within the graph state formalism.
A short summary of the results and open question are provided in Section~\ref{section:conclusion}.

\emph{Notation:} Throughout this paper, if the dimension of a system is denoted by $p$, it is meant to be a prime number. If we use $d$ for the dimension of a system, no constraints are imposed.

\section{Qudit Graph States}
\label{section:quditgraphs}

\subsection{Generalized Pauli Operators}
The generalized Pauli operators \cite{Patera1988, Gottesman2001, Bartlett2002, Keet2010} for qudits of dimension $d$ are defined as
\begin{align}
\label{eq:genZ}
 Z \ket{k} &= \omega^k \ket{k},\\
\label{eq:genX}
 X \ket{k} &= \ket{k+1},
\end{align}
where $\omega = e^{2\pi i/d}$. Controlled gates are generalized straightforward, with the controlled-$Z$ operator between qudit $i$ and $j$ being
\begin{equation}
 \label{eq:genCZ}
 \CZ_{ij} = \sum_{k=0}^{d-1} \ketbraind{k}{k}{i} \otimes Z_j^k
 = \sum_{k,l=0}^{d-1} \omega^{kl} \ketbraind{k}{k}{i} \otimes \ketbraind{l}{l}{j}
\end{equation}
It is easily seen that $Z^d = X^d = \CZ^d = \Id$. Furthermore we have the commutation relation $ZX = \omega XZ$. The Fourier gate
\begin{equation}
 F = \frac{1}{\sqrt{d}} \sum_{k=0}^{d-1} \omega^{kl} \ket{k}\bra{l},
\end{equation}
the generalization of the Hadamard gate, transforms between the $Z$-eigenbasis $\ket{k}$, and the $X$-eigenbasis $\ket{\bar{k}}$,
\begin{equation}
 \ket{\bar{k}} = F^\dagger \ket{k} = \frac{1}{\sqrt{d}} \sum_{l=0}^{d-1} \omega^{-kl} \ket{l}.
\end{equation}

\subsection{Graph States}
We are now ready to define graph states for $n$ qudits of dimension $p$, where $p$ is a prime number. The qudits are graphically represented by \emph{vertices} $\cV = \{v_i\}$, which are connected by \emph{edges} $\cE = \{e_{ij} = \{v_i,v_j\}\}$. Each edge is assigned a \emph{weight} $A_{ij} \in \ZZ_p$, where weight zero is equivalent to no edge. The weights $A_{ij}$ form the symmetric $n \times n$ \emph{adjacency matrix} with $A_{ii} = 0$ that captures all the relevant information about the graph.

\begin{definition}
For a given graph \gG with $n$ vertices and adjacency matrix $A \in \ZZ_p^{n\times n}$, where $p$ is prime, we define the corresponding graph state $\ket{G} \in \HH^{\otimes n}$, $\HH \cong \CC^p$ as
\begin{equation}
 \label{eq:graph}
 \ket{G} = \prod_{i > j} \CZ_{ij}^{A_{ij}} \ket{\bar{0}}^{\otimes n}.
\end{equation}
We further define a \emph{labeled graph states} by attaching an additional label $\vv z = (z_1, \ldots, z_n) \in \ZZ_p^n$ to the graph state $\ket{G}$ as
\begin{equation}
 \label{eq:labeledGraph}
 \ket{G_{\vv z}} = Z^{\vv z} \ket{G},
\end{equation}
Here and in the following we use the notation
\begin{equation}
 Z^{\vv z} = Z^{z_1} \otimes Z^{z_2} \otimes \cdots \otimes Z^{z_n}.
\end{equation}
\end{definition}

A graph state can be constructed by a quantum circuit that first prepares all systems in the $\ket{\bar{0}}$ state, and then applies pairwise controlled-Z gates between the systems according to the entries of the adjacency matrix.

\subsection{Stabilizer States}
Stabilizer states have first been introduced for qubits \cite{GottesmanPhD} and later generalized to qudits \cite{Ashikhmin2001, Ketkar2006}. The connection to qudit graph states has been made in Ref.~\cite{Schlingemann2002, Bahramgiri2006}. The \emph{Pauli Group}, the group that is generated by the $X$ and $Z$ operators for qubits is defined as
\begin{equation}
 \label{eq:PauliGroup}
\PG = \{\alpha X^a Z^b; a,b \in \ZZ_2 \},
\end{equation}
with $\alpha \in \{1, -1, i, -i\}$, and its generalization for the qudit Pauli operators of \cref{eq:genZ,eq:genX} is
\begin{equation}
 \label{eq:genPauliGroup}
\PG = \{\omega^c X^a Z^b; a,b,c \in \ZZ_p \},
\end{equation}
with $\omega = e^{2\pi i/p}$. The Pauli group over $n$ qudits is the $n$-fold tensor product of \PG and is denoted $\PG_n$.

The \emph{stabilizer code} is defined as the common eigenspace for eigenvalue one of a subgroup $S$ of $\PG_n$. The stabilizer code is non-trivial if $S$ is abelian and does not contain any scalar multiples of the identity, except for $\Id$ itself. \cite{Ketkar2006, Bahramgiri2006}. Given such a subgroup and a minimal set of generators, $g_i = \omega^{c_i} X^{\vv{a_i}} Z^{\vv{b_i}}$, for the group, $S = \langle g_1, \dots, g_k \rangle$, the \emph{generator matrix} is defined as
\begin{equation}
 \label{eq:stabilizergeneratormatrix}
 M = \left(
 \begin{array}{c|c}
  \vv{a_1} & \vv{b_1}\\
  \vdots & \vdots\\
  \vv{a_k} & \vv{b_k}
 \end{array}\right).
\end{equation}
The stabilizer code does not depend on the scalar coefficients $\omega^{c_i}$, and is thus fully specified by the generator matrix $M$. The fact that $S$ is abelian translates to $\vv{a_i}\cdot \vv{b_j} - \vv{b_i}\cdot \vv{a_j} = 0$ for two different rows of $M$. It has been shown in Ref.~\cite{Bahramgiri2006} that a stabilizer group $S$ with $k$ generators corresponds to a stabilizer code of dimension $n-k$. Thus if the minimal set of generators for $S$ is of size $n$, then the stabilizer code only contains one state, the \emph{stabilizer state} to the generator matrix $M$. 

A special class of stabilizer states are the above introduced graph states. Given the adjacency matrix $A$, a minimal set of generators for the stabilizer group is given by
\begin{equation}
 \label{eq:graphstabilizers}
 g_i = X_i\prod_j Z_j^{A_{ij}}.
\end{equation}
Here the indices labels on which qudit the operator act. This means the generator matrix is simply given by
\begin{equation}
 M = \left( \Id | A \right)
\end{equation}

The Clifford group, the group of operators that maps the Pauli group onto itself, can also be generalized to qudits (for details on the generalized Clifford group, see Ref.~\cite{Hostens2005}). The local Clifford group for a system of $n$ qudits is the $n$-fold tensor product of the Clifford group. The following lemma, which shows when two states can be transformed into each other by an element of the local Clifford group, is proved in Ref.~\cite{Bahramgiri2006}.

\begin{lemma}[Lemma 6 of Ref.~\cite{Bahramgiri2006}]
  \label{lemma:equivalentCliffordGroup}
 Two stabilizer states with generator matrices $A$, $B$ are equivalent under the action of the local Clifford group, if and only if there exist invertible matrices $U$ and $Y$, such that $B = UAY$, and $Y$ has the form
 \begin{equation}
  Y = \left( \begin{array}{cc}
              E & F\\
              E' & F'
             \end{array}\right),
 \end{equation}
  where 
  \begin{align}
   E = \diag (e_1,\ldots, e_n), \qquad & F= \diag (f_1,\ldots, f_n)\\
   E' = \diag (e'_1,\ldots, e'_n), \qquad & F'= \diag (f'_1,\ldots, f'_n),
  \end{align}
  and $e_i f'_i - f_i e'_i = 1$ for all $i$.
\end{lemma}
It has been further shown that every stabilizer state is equivalent to a graph state under the action of the local Clifford group \cite{Bahramgiri2006, Schlingemann2002, Grassl2002}. Thus if we want to consider possible entanglement properties of stabilizer states, it suffices to consider graph states, since for any stabilizer state there exists a graph state with the same entanglement properties.

\section{Entanglement in Graph States}
\label{section:entanglement}
Now that we have introduced qudit graph states, the next question is, given a certain graph state, described by the adjacency matrix $A$ for $n$ qudits, how to determine the entanglement of the associated quantum state. Specifically, we are interested in the entanglement between bipartitions of the $n$ parties. If all these bipartitions are maximally entangled, the state is an absolutely maximally entangled state.

We present two different methods for checking the entanglement between bipartitions. The first uses the fact that the entanglement can be determined just by looking at the graph, if it is in the right form. The problem in this method is to bring the graph state into the right form for any bipartition. This is generally not so easy and thus we also present a second method that is computationally more helpful to actually determine the bipartite entanglement in graph states.

\subsection{Graphical Representation}
\begin{figure}[htb]
 \centering
 \hfill
 \subfigure[]{
 \includegraphics[width=.25\textwidth]{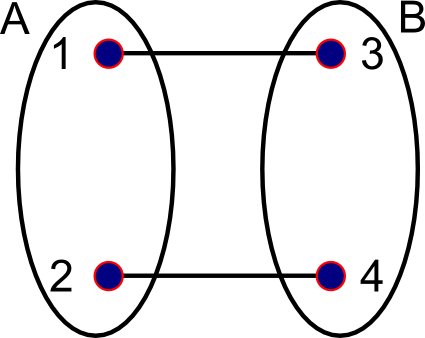}
 \label{fig:4qudits1}
 }
 \hfill
 \subfigure[]{
 \includegraphics[width=.25\textwidth]{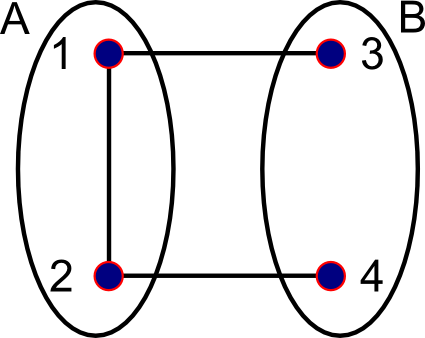}
 \label{fig:4qudits2}
 }
 \hfill
 \subfigure[]{
 \includegraphics[width=.25\textwidth]{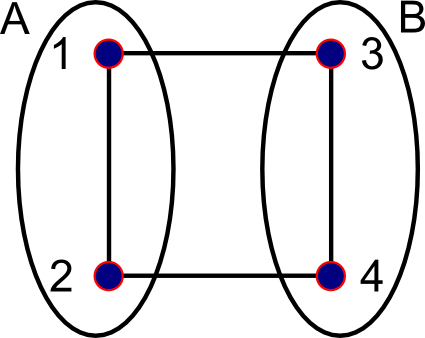}
 \label{fig:4qudits3}
 }
 \hfill
 \caption{Graph states for four qudits with maximal entanglement between the sets $A$ and $B$. The graph in (a) is the simplest graph that shows maximal entanglement between the sets $A$ and $B$. Adding edges within each set is only a local operation with regard to that bipartition and thus does change the entanglement properties between the sets. Thus all the shown graphs have the same amount of entanglement between $A$ and $B$.}
\end{figure}

Recall that an edge of the graph represents the application of a controlled-Z gate. If a controlled-Z gate is applied between two qudits in the $\ket{\bar{0}}$ state, they are maximally entangled. We say they share 1 ``\emph{edit}'' of entanglement.
For $n$ qudits, divided into two sets $A$ and $B$, the maximal amount of entanglement between the two sets is $\min(|A|,|B|)$ edits. This can, for instance, be achieved by preparing each qudit in the $\ket{\bar{0}}$ state, and then applying controlled-Z gates between the qudits, such that each party of the smaller set is connected to a different party in the larger set. An example of a resulting graph for four qudits is depicted in Figure~\ref{fig:4qudits1}. 

Applying the controlled-Z gate up to $p-1$ times between two qudits also creates 1 edit of entanglement, thus we may assign any non-zero weight to the connecting edges, without changing the maximal entanglement. Furthermore, applying local unitary operations within each set after the entanglement between the sets has been created will not change the entanglement between these sets. Thus we may add as many edges with arbitrary weight as we like within each set, and the sets $A=\{1,2\}$ and $B=\{3,4\}$ will still remain maximally entangled. This is demonstrated in the graphs in Figures~\ref{fig:4qudits2} and \ref{fig:4qudits3}, which are both still maximally entangled for the bipartition into the sets $A$ and $B$.

Checking the entanglement in these graphs between the sets $A$ and $B$ is easy, because we specifically constructed the state that way. The entanglement for the bipartition $\{1,3\}/\{2,4\}$ is also obvious, it is 0, 1, and 2 edits for the states in Figures~\ref{fig:4qudits1}, \ref{fig:4qudits2}, and \ref{fig:4qudits3}, respectively. 
However, the entanglement for a different bipartition, for example between $C=\{1,4\}$ and $D=\{2,3\}$ in the graph of Figure~\ref{fig:4qudits3} is not immediately obvious. To determine the entanglement, we need a graph in which each party in $C$ is connected to at most one party in $D$ and vice versa. Then counting the number of connecting edges gives the number of edits shared between $C$ and $D$. Changes allowed on the graph are the ones that don't change the entanglement properties of the graph, like the ones achieved by local Clifford operations as described in Lemma~\ref{lemma:equivalentCliffordGroup}. For graph states this lemma can be restated as operations on the graph \cite{Bahramgiri2006}

\begin{theorem}[Theorem 5 of Ref.~\cite{Bahramgiri2006})]
Two graph states are equivalent under local Clifford operations if and only if one can be obtained from the other by a sequence of the two graph operations on a vertex $v$
\begin{itemize}
 \item[$\circ_b v$] The weight of each edge connected to the vertex $v$ is multiplied by $b$, where $0\neq b \in \ZZ_p$. 
 \item[$\ast_a v$] For $a \in \ZZ_p$, the entries of the adjacency matrix are transformed as $A_{jk} \rightarrow A_{jk} + aA_{vj} A_{vk}$ for $j\neq k$. 
\end{itemize}
\end{theorem}
A graphical representation of these operations, is given in Figures 1 and 2 of Ref.~\cite{Bahramgiri2006}. For qubits the $\circ$ operation is always the identity, and the $\ast$ operation for $a=1$ is known as the \emph{local complementation}. Returning to the question what the entanglement between sets $C=\{1,4\}$ and $D=\{2,3\}$ is in the graph of Figure~\ref{fig:4qudits3}, we can see that by applying the operations $(\ast_1 1,\ \ast_1 3,\ \ast_1 4)$, the graph is in fact local Clifford equivalent to the graph of Figure~\ref{fig:4qudits2} with vertices 3 and 4 interchanged. Hence it shares only 1 edit of entanglement for the bipartition into sets $C=\{1,4\}$ and $D=\{2,3\}$, and therefore is not absolutely maximally entangled. Examples of states where this method confirms absolutely maximal entanglement will be given in Section~\ref{section:AMEstates}.

\subsection{Efficient Method}
\label{sec:EntanglementInd}
While the above presented method to determine bipartite entanglement is very intuitive, it is generally not easy to find the right graph operations to bring the graph into the right form for a given bipartition. Thus we will present a second method that makes it computationally relatively easy to check the bipartite entanglement for a given graph for an arbitrary bipartition. We will make use of the following notations

\begin{definition}
 Let $\ket{G}$ be a graph state shared between a set of parties $P$. Then, for $K\subset P$, we define the truncated graph state $\ket{G^{\backslash K}}$, shared by $P\backslash K$, as the state that is represented by the graph $G$ with the vertices in $K$ and all edges that are connected to the parties in $K$ removed.
\end{definition}

\begin{definition}
 For an $n\times n$ adjacency matrix $A$, we denote the $i$th row of the matrix by $A_i$, so $A_i = (A_{i1}, \ldots A_{in})$. Further for $K = \{ k_1, k_2, \ldots, k_m \}$, with $k_1, \ldots, k_m$ between 1 and $n$, we define $A_i \backslash K$ to be the vector $A_i$ with the entries $\{A_{ik_1}, \ldots, A_{ik_m}\}$ removed. For instance, 
 \begin{equation}
  A_i\backslash \{2,6\} = (A_{i1}, A_{i3}, A_{i4}, A_{i5}, A_{i7}, \ldots, A_{in}).
 \end{equation}
\end{definition}

First note that a $Z$-measurement\footnote{$Z$ is not technically an observable, what we mean by a $Z$-measurement is a projection onto the eigenstates with (complex) eigenvalues $\omega^k$. For simplicity we then call the measurement result $k$.} on the $k$th qudit of the graph
\begin{align}
 \ket{G}
  &= \prod_{i > j} \CZ_{ij}^{A_{ij}} \ket{\bar{0}}^{\otimes n}\\
  &=  
    \prod_{l \neq k} \sum_{m=0}^{p-1} \ketbraind{m}{m}{k} \otimes Z_l^{mA_{kl}}
    \prod_{\substack{i>j \\ i,j\neq k}} \CZ_{ij}^{A_{ij}} 
    \ket{\bar{0}}^{\otimes n}
  ,
\end{align}
with measurement outcome $a$ gives
\begin{align}
  \braketind{k}{a|G}{} 
  &= \frac{1}{\sqrt{p}} \prod_{l \neq k} Z_l^{aA_{kl}}
    \prod_{\substack{i>j \\ i,j\neq k}} \CZ_{ij}^{A_{ij}} 
    \ket{\bar{0}}^{\otimes n-1}\\
   &= \frac{1}{\sqrt{p}}  \ket{G^{\backslash \{k\}}_{a A_k\backslash\{k\}}}.
\end{align}
So this is a labeled graph state for the remaining $n-1$ qudit, with the label given by $A_k\backslash\{k\}$ with each component multiplied by the measurement outcome $a$.
All measurement outcomes are equally likely, and given that $A_k\backslash\{k\} \neq 0$, meaning that the $k$th qudit in $\ket{G}$ is connected  by at least one edge, the label is different for each possible measurement outcome. Since labeled graph states with different labels are orthogonal, the measurement outcome can be deduced from $\ket{G^{\backslash \{k\}}_{a A_k\backslash\{k\}}}$. Hence qudit $k$ is maximally entangled with the other $n-1$ qudits in $\ket{G}$.

Similarly, if $Z$-measurements are performed on $m$ qudits $K=\{k_1,\dots,k_m\}$, with measurement outcomes $\{a_{1},\ldots,a_{m}\}$, the resulting state is
\begin{align}
  \braketind{k_1,\ldots,k_m}{a_{1},\ldots,a_{m}|G}{} 
   &= \frac{1}{\sqrt{p^m}}  \ket{G^{\backslash K}_{\sum_i^m a_{i} A_{k_i}\backslash K}}.
\end{align}
This again is a labeled graph state with the measured qudits and associated edges removed, and the $Z$ operations applied for each measurement independently, because $Z$ measurements and $Z$ operators commute. Note that $Z$ operations on qudits in $K$ only contribute as a global phase, which we have omitted.

If the label is different for each different possible combination of measurement outcomes $\{a_{1},\ldots,a_{m}\}$, the resulting labeled graph states are all orthogonal and the remaining parties can determine the measurement outcome. Thus the parties in $K$ share $m$ edits of entanglement with the other $n-m$ parties. The labels are all different if and only if the $m$ vectors $A_{k_i}\backslash\{k_1,\dots,k_m\}$ are linearly independent in $\ZZ_p^{n-m}$. Thus we have the following theorem:

\begin{theorem}
 \label{theorem:independentVectors}
 A graph state with adjacency matrix $A$ is absolutely maximally entangled, if and only if for all sets $K=\{k_1,\dots,k_m\}$ of size $m=\floor{\frac{n}{2}}$, the vectors $A_{k_i}\backslash K$ are linearly independent in $\ZZ_p^{n-m}$. Here $A_{k_i}\backslash K$ denotes the $k_i\tth$ row of the adjacency matrix with the entries $\{A_{k_ik_1}, \ldots, A_{k_ik_m}\}$ removed.
\end{theorem}

As a concrete example, we take a look at the graph of Figure~\ref{fig:4qudits3} again and use this method to determine if it is absolutely maximally entangled. For the bipartition into the sets $K=\{1,2\}$ and $L=\{3,4\}$, we get the two vectors $A_1\backslash\{1,2\} = (1,0)$ and $A_2\backslash\{1,2\} = (0,1)$. These are independent and thus we have maximal entanglement between the sets $K$ and $L$. We get the same vectors for the bipartition $\{1,3\}/\{2,4\}$, so we also have maximally entanglement there. However, for the bipartition into $C=\{1,4\}$ and $D=\{2,3\}$, we get the vectors $A_1\backslash\{1,4\} = (1,1)$ and $A_4\backslash\{1,4\} = (1,1)$. These are not independent and thus we do not have maximal entanglement for this bipartition. Since their is only one independent vector, this bipartition shares 1 edit of entanglement.

\section{Absolutely Maximally Entangled Graph States}
\label{section:AMEstates}

\subsection{Qubits, Qutrits, and Beyond}
\begin{figure}[htb]
 \centering
 \hfill
 \subfigure[\AME{2,2}]{
 \raisebox{0.7cm}{\includegraphics[width=.2\textwidth]{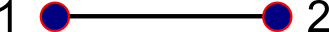}}
 \label{fig:AME22}
 }
 \hfill
 \subfigure[\AME{3,2}]{
 \raisebox{0.5cm}{\includegraphics[width=.25\textwidth]{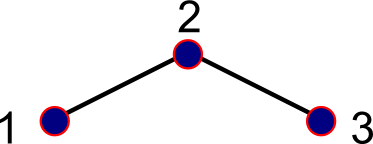}}
 \label{fig:AME32}
 }
 \hfill
 \subfigure[\AME{5,2}]{
 \includegraphics[width=.25\textwidth]{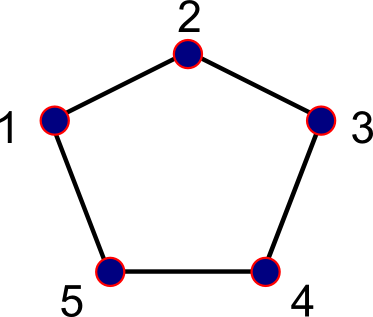}
 \label{fig:AME52}
 }
 \hfill\\
 \hfill
 \subfigure[\AME{6,2}]{
 \includegraphics[width=.3\textwidth]{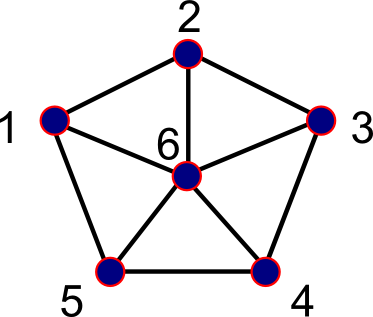}
 \label{fig:AME62wheel}
 }
 \hfill
 \subfigure[\AME{6,2}]{
 \includegraphics[width=.25\textwidth]{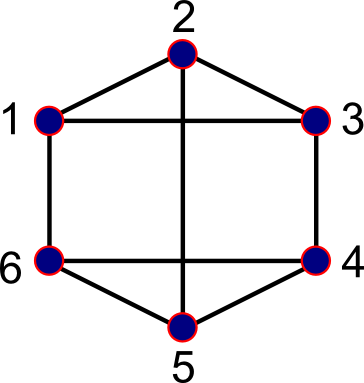}
 \label{fig:AME62}
 }
 \hfill\hfill
 \caption{Absolutely maximally entangled qubit graph states exist for two, three, five and six systems. The two qubit state is locally equivalent to an EPR pair and the three qubit state to a GHZ state. The five qubit AME state finds application in the five qubit code and quantum secret sharing. The six qubit state in Figure~\ref{fig:AME62wheel} emphasizes the connection to the five qubit state, while the locally equivalent state of Figure~\ref{fig:AME62} nicely demonstrates the maximal entanglement.}
 \label{fig:qubitAMEstates}
\end{figure}

It is known \cite{Helwig2012, Helwig2013} that for qubits there exist absolutely maximally entangled states for 2, 3, 5 and 6 qubits. In all these cases, we can also find absolutely maximally entangled graph states. They are given in Figure~\ref{fig:qubitAMEstates}. The ones for two and three qubits are the well Einstein-Podolsky-Rosen (EPR) pair and the Greenberger-Horne-Zeilinger (GHZ) state, respectively.
The AME states in Figures~\ref{fig:AME52} and \ref{fig:AME62wheel} for five and six qubits can be used for quantum secret sharing protocols \cite{Markham2008}, as will also be discussed in the next section.
We also included a second graph for six qubits in Figure~\ref{fig:AME62}, which illustrates the maximal entanglement for the bipartition $\{1,2,3\} / \{4,5,6\}$, when using the graphical method to check for maximal entanglement. This is also the representation with the least number of edges and it is locally Clifford equivalent (related by a $\ast_1 v$ operation) to the one in Figure~\ref{fig:AME62wheel}.

For four and more than eight qubits no AME states exists. For seven qubits no AME states are known, and and exhaustive search of all seven qubit graph states showed that no seven qubit AME graph state exists. Increasing the system dimension, however, can can help us to find AME states for scenarios where no qubit AME states exist. The reason for that is that with higher system dimension $p$, the graph can have $p-1$ different types of weighted edges. This exponential growth of possible graphs results in a greater variety of entanglement properties, which allows to construct more graphs that are absolutely maximally entangled.

\begin{figure}[htb]
 \centering
 \subfigure[\AME{4,3}]{
 \includegraphics[width=.25\textwidth]{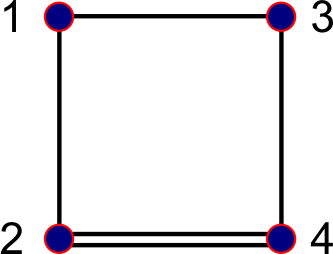}
 \label{fig:AME43a}
 }
 \raisebox{1cm}{$\xrightarrow{\ast_1 1,\ \ast_1 3,\ \ast_1 2}$}
 \subfigure[\AME{4,3}]{
 \includegraphics[width=.25\textwidth]{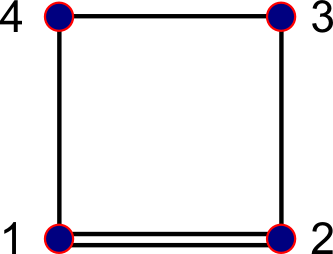}
 \label{fig:AME43b}
 }
 \hfill
 \caption{Absolutely maximally entangled graph states for four qutrits. The first one demonstrates the maximal entanglement for the bipartitions $\{1,2\}/\{3,4\}$ and $\{1,3\}/\{2,4\}$. The second graph is locally Clifford equivalent to the first and shows the maximal entanglement for the bipartition $\{1,4\}/\{2,3\}$.}
 \label{fig:AME43state}
\end{figure}

\begin{figure}[htb]
 \centering
 \includegraphics[width=.25\textwidth]{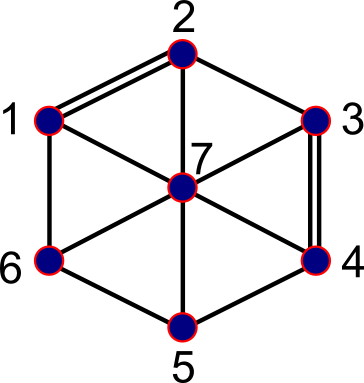}
 \caption{\AME{7,3} graph state. The use of double edges allows us to find an AME graph state for seven qutrits, while no such graph state exists for qubits.}
 \label{fig:AME73state}
\end{figure}

The first time we see that is for four qudits. If we only consider graphs with edges of weight one, which are the only ones available for qubits, the graph with the most amount of entanglement we can construct is the one in Figure~\ref{fig:4qudits3}. This graph is maximally entangled for two of the three possible bipartition, but not for the third as discussed in the last section. Hence we have to consider graphs that have edges with higher weights, for instance the graph state shown in Figure~\ref{fig:AME43a}, where we have assigned the weight 2 to one of the four edges. This graph is obviously still maximally entangled for the $\{1,2\}/\{3,4\}$ and $\{1,3\}/\{2,4\}$ bipartitions. To check the entanglement for the $C=\{1,4\}/D=\{2,3\}$ bipartition with the graphical method, we have to perform the operations $(\ast_1 1,\ \ast_1 3,\ \ast_1 2)$ to obtain the graph shown in Figure~\ref{fig:AME43b}, from which we see that the $C/D$ bipartition is also maximally entangled. Likewise, we could have 
considered the two vectors $A_1\backslash\{1,4\}=(1,1)$ and $A_4\backslash\{1,4\}=(2,1)$, to see that they are linearly independent in $\ZZ_3$. Hence we have just confirmed that this graph is maximally entangled for four qutrits.

By doing a computer search over highly entangled seven qutrit states, the efficient method of Section~\ref{sec:EntanglementInd} for checking bipartite entanglement in graph states allowed us to find an \AME{7,3} graph state. It is displayed in Figure~\ref{fig:AME73state}.

These examples nicely illustrate that by increasing the system dimension, more AME graph states can be found due to the increased number of graph configurations.

\begin{figure}[htb]
 \centering
 \includegraphics[width=.25\textwidth]{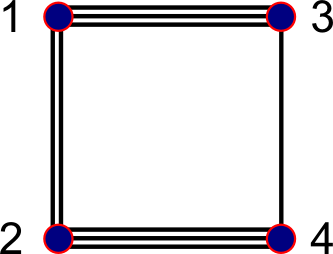}
 \caption{An AME state for one dimension is not neccessarily a graph state for a higher dimension. For instance, this graph states is absolutely maximally entangled for four qudits of dimension 5, but not for qudits of dimension 7.}
 \label{fig:AME45state}
\end{figure}

Another nice property of the AME graph states is that the same graph state often works for more than one dimension. For instance the qubit graph states of Figure~\ref{fig:qubitAMEstates} are AME states for any prime dimension, because if a set of vectors is independent in $\ZZ_2$, they are also independent in $\ZZ_p$. Also the graph state in Figure~\ref{fig:AME43a} is an AME state for any prime dimension $p\geq 3$, because the vectors $(1,1)$ and $(2,1)$ are independent in all $\ZZ_p^2$ for $p\geq 3$. However, it is not always the case that AME graph states generalize to all higher prime dimensions. A counter-example is given in Figure~\ref{fig:AME45state}, which shows a graph state that is absolutely maximally entangled for $p=5$, but not for $p=7$ because for the bipartition $\{1,4\}/\{2,3\}$, we have to check the two vectors $(2,3)$ and $(3,1)$ for independence, and these two vectors are independent in $\ZZ_5^2$, but not in $\ZZ_7^2$.

\subsection{AME Graph States from Classical Codes}
By now we have seen AME graph states for system with up to seven parties. In Ref.~\cite{Helwig2013}, it was shown that AME states can be constructed from classical error correction codes and that linear codes of the required form exist for any number of parties if the dimension of the systems is chosen appropriately. A linear code $\Ccal$, which encodes $k$ \emph{dits} of information into $n$ dits, is described by a \emph{generator matrix} $G: \ZZ_p^k \rightarrow \ZZ_p^n$ such that the codewords $c \in \Ccal$ are given by $G \vv x$ for $\vv x \in \ZZ_p^k$. An equivalent description of a linear code as the kernel of the \emph{parity check} matrix $H$. For every linear code $\Ccal$ one can define a parity check matrix $H: \ZZ_p^n \rightarrow \ZZ_p^{n-k}$ such that $c \in \Ccal$ if and only if $H c = 0$. From $HGx = 0$, it follows that the rows of $H$ are orthogonal to the columns of $G$.

The \emph{Hamming distance} between two codewords is defined as the number of positions at which the codewords differ. The \emph{minimal distance} $\delta$ of a code is the minimum Hamming distance between any two codewords. The larger $\delta$, the more robust the encoding is against errors. The minimal distance is bounded by the Singleton bound, $\delta \leq n-k+1$ \cite{Singleton1964, MacWilliams1977}. Codes that satisfy the Singleton bound are called \emph{maximum distance separable (MDS)} codes. 

A MDS code with the properties $n=2k$, $\delta=k+1$ can be used to construct an AME state. The AME state is then given by \cite{Helwig2013}.
\begin{equation}
 \ket{AME} = \frac{1}{\sqrt{d^k}} \sum_{\vv x\in \ZZ_p^k} \ket{G \vv x}.
\end{equation}
Note that
\begin{align}
 X^{G\vv y} \ket{AME} &= \frac{1}{\sqrt{d^k}} \sum_{\vv x\in \ZZ_p^k} \ket{G\vv x + G\vv y}\\
 &= \frac{1}{\sqrt{d^k}} \sum_{\vv x\in \ZZ_p^k} \ket{G\vv x}\\
 &= \ket{AME},
\end{align}
where we have used that $\Ccal$ is a linear code and the sum goes over all codewords of the code. Thus adding the same codeword to all other codewords is just a relabeling of the terms in the sum. Thus $X^{G \vv y}$ is a stabilizer to the AME state for all $\vv y \in \ZZ_p^k$. Another set of stabilizers can be constructed from the $Z$-Operators. The action of a tensor product of $Z$-operators on the AME  state is given by
\begin{equation}
 Z^{\vv y} \ket{AME} =  
  \frac{1}{\sqrt{d^k}} \sum_{\vv x\in \ZZ_p^k} \omega ^{\vv y^T G \vv x} \ket{G\vv x}.
\end{equation}
Thus $Z^{\vv y}$ is a stabilizer for the AME state if $\vv y$ is a linear combination of rows of the parity check matrix $H$, $\vv y^T = \vv z^T H$. This gives us a full set of stabilizers that we can describe by the generator matrix
\begin{equation}
 M = \left(
 \begin{array}{c|c}
  G^T & 0\\
  0 & H
 \end{array}\right).
\end{equation}
It is easy to see that all the generators are independent, since the columns of $G$ and rows of $H$ are linearly independent. They are also abelian as they satisfy $\vv{a_i} \cdot \vv{b_j} - \vv{b_i} \cdot \vv{a_j} = 0$ because the rows of $H$ are orthogonal to the columns of $G$. Thus $M$ is a proper generator matrix to the stabilizer state $\ket{AME}$. Given the generator matrix $M$, the AME state can now be transformed into a graph state by local Clifford operations that change the generator matrix according to Lemma~\ref{lemma:equivalentCliffordGroup} \cite{Bahramgiri2006}.

The whole procedure of constructing an AME graph state from an MDS code is illustrated in the following example for the $[4,2,3]_3$ ternary Hamming code that results in an \AME{4,3} graph state.

\begin{figure}[htb]
 \centering
 \includegraphics[width=.25\textwidth]{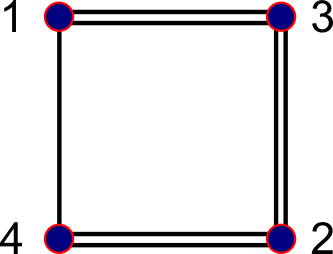}
 \caption{\AME{4,3} graph state constructed from the $[4,2,3]_3$ ternary Hamming code}
 \label{fig:AME43fromMDS}
\end{figure}

\begin{example}
 The generator matrix for the $[4,2,3]_3$ ternary Hamming code $\Ccal$ is given by
 \begin{equation}
  G = \left(
  \begin{array}{cccc}
   1&0\\
   0&1\\
   1&1\\
   2&1
  \end{array}\right),
 \end{equation}
  and the parity check matrix by $H = G^T$ ($\Ccal$ is a self-dual code). Thus the generator matrix for the AME  state $\ket{AME} = \frac{1}{3} \sum_{c\in \Ccal} \ket{c}$ is
  \begin{equation}
   M = \left(
   \begin{array}{cccc|cccc}
    1&0&1&2 & 0&0&0&0\\
    0&1&1&1 & 0&0&0&0\\
    0&0&0&0 & 1&0&1&2\\
    0&0&0&0 & 0&1&1&1
   \end{array}\right)
  \end{equation}
 Now we have to choose the matrices $U$ and $Y$ of Lemma~\ref{lemma:equivalentCliffordGroup} such that $UMY$ is the identity matrix in the first block. For that note that by choosing $f_i=0$ and $e_i=f'_i=1$, the condition for $Y$ is satisfied for arbitrary $e'_i$. The effect of the value $e'_i$ is to add the $i$th column of the second block to the $i$th column of the first block. We want to choose them such that the first block has full rank, which is accomplished by $e_1 = e_2 = 0$ and $e_3=e_4=1$. This transforms the generator matrix to
 \begin{equation}
  M \rightarrow MY = 
  \left(
   \begin{array}{cccc|cccc}
    1&0&1&2 & 0&0&0&0\\
    0&1&1&1 & 0&0&0&0\\
    0&0&1&2 & 1&0&1&2\\
    0&0&1&1 & 0&1&1&1
   \end{array}\right).
 \end{equation}
 Then we have to choose $U$ such that it transforms the first block into the identity. This is achieved by
 \begin{equation}
  U=
  \left(
  \begin{array}{cccc}
   1&0&2&0\\
   0&1&0&2\\
   0&0&2&2\\
   0&0&1&2
  \end{array}\right),
 \end{equation}
 which results in the generator matrix
 \begin{equation}
  M \rightarrow UMY = 
  \left(
   \begin{array}{cccc|cccc}
    1&0&0&0 & 2&0&2&1\\
    0&1&0&0 & 0&2&2&2\\
    0&0&1&0 & 2&2&1&0\\
    0&0&0&1 & 1&2&0&1
   \end{array}\right).
 \end{equation}
  This generator matrix has the desired form except for the entries on the diagonal of the second block, which may be transformed to zero by an additional application of a $Y$ matrix with $e'_i=0$, $e_i=f'_i=0$, and $(e_1,e_2,e_3,e_4) = (1,1,2,2)$. Thus we arrived at the graph state shown in Figure~\ref{fig:AME43fromMDS}, which is an absolutely maximally entangled graph state for four qutrits.
\end{example}

Notice that the procedure of constructing a stabilizer state from classical codes is reminiscent of the construction of Calderbank-Shor-Steane (CSS) codes \cite{Calderbank1996, Steane1996}. In fact we may interpret the AME states that are constructed in this form as one-dimensional generalized CSS codes $\mathrm{CSS}_p(\Ccal,\Ccal)$.

\subsection{Non-prime dimensions}
So far all we considered were scenarios where the parties shared systems of prime dimension. This was because although the initial definition of graph states in terms of controlled-$Z$ gates applied to qudits in the $\ket{\bar 0}$ state works for any dimension, the following treatment in terms of stabilizers does not. This includes, in particular, the methods we derived to check the entanglement of graph states in Section~\ref{section:entanglement}. In reality, however, we might have to deal with systems that are not of prime dimension, so how can we still describe them while taking advantage of the tools the graph state formalism provides us with for prime dimensions?

The answer is, we take the prime factorization for the system dimension $d=p_1 \cdot p_2 \cdots p_m$ and look for AME states for $p_1, \ldots p_m$ independently, and if we have an AME state for each of the prime factors, then we can just construct an AME state for $d$ by taking the tensor product of the $m$ AME states and assigning one qudit of each AME state to each of the parties. In this way, we can, for instance, construct AME states for any dimension for the number of parties $n= 2, 3, 5, 6$, since the known qubit AME graph states work for any dimensions. Likewise, we can construct a four qudit AME state for any uneven dimension, since the AME graph state of Figure~\ref{fig:AME43a} generalizes to all prime dimensions $p\geq 3$.

\begin{figure}[htb]
 \centering
 \includegraphics[width=.35\textwidth]{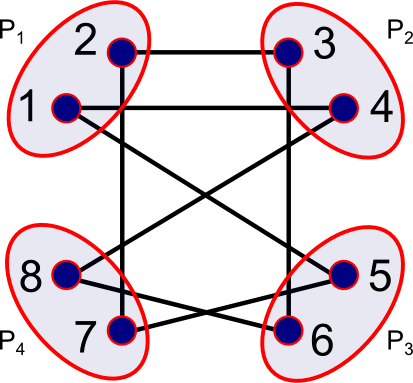}
 \caption{By grouping qudits together, we can construct AME graph states for non-prime dimensions. This figure shows eight qudits that are grouped into four 4-dimensional systems to form an \AME{4,4} graph state. This, however, is not an \AME{8,2} state if each qudit is regarded as a single party.}
 \label{fig:AME44state}
\end{figure}

Furthermore, if two or more of the prime factors are the same, for instance for $d=4=2\cdot 2$, we may apply controlled-$Z$ operations between one qubit of one party and either qubit of the other parties. This is best illustrated in an example. Imagine we want to find an \AME{4,4} state. It is not possible to simply take two \AME{4,2} states, because they do not exist. We can, however, consider the each 4-dimensional systems as consisting of two qubits and construct the graph state shown in Figure~\ref{fig:AME44state} for eight qubits. This state is maximally entangled with 4 ebits (=2 edits) of entanglement for the bipartitions $\{P_1,P_2\} / \{P_3, P_4\}$, $\{P_1,P_3\} / \{P_2, P_4\}$ and $\{P_1,P_4\} / \{P_2, P_3\}$. Thus this graph state describes an \AME{4,4} state. Note that this state is generally not maximally entangled for bipartitions where we split up the two qubits belonging to one party, and is thus not an \AME{8,2} state (which does not exist).

\section{Quantum Secret Sharing}
\label{section:QSS}
One application for AME states is to construct quantum secret sharing (QSS) protocols \cite{Cleve1999, Gottesman2000, Helwig2012, Helwig2013}. In a quantum secret sharing protocol, a secret is encoded into a quantum state shared between $n$ players $P$ such that certain subsets of $P$, the \emph{authorized} sets, are able to recover the secret by performing joint quantum operations, while other subsets, the \emph{forbidden} sets, are not able to gain any information about the secret. In some secret sharing protocols, for instance in threshold QSS schemes \cite{Cleve1999, Gottesman2000}, any subset is either authorized or forbidden. However, in general there may also exist a third category, the \emph{intermediate} sets, which are subsets of players that are not able to recover the full secret, but are able to gain some information about the secret. 

Describing quantum secret sharing protocols with the help of graph states has already been studied before for qubit \cite{Markham2008} and qudit graph states of prime dimension \cite{Keet2010}. In these papers it was shown that threshold quantum secret sharing schemes can be constructed from the graph state shown in Figure~\ref{fig:AME62wheel} for 6 qudits of arbitrary prime dimension, and for the graph state shown in Figure~\ref{fig:AME43a} for four qudits of prime dimension $p\geq 3$. However, the question which graph states are generally suitable for quantum secret sharing remained an open question.

\subsection{Threshold QSS Schemes}
In a pure state $((m,2m-1))$ threshold QSS scheme, the secret is encoded into a pure state that is distributed among an odd number of players $P=\{1,\ldots,2m-1\}$ such that a subset $B\subset P$ of players is authorized if and only if the set contains more than half the players, $|B| \geq m$. Furthermore, a subset $B \subset P$ of players with less than $m$ players is always a forbidden set.

It was already shown, that there exists a one-to-one correspondence between pure state \QSS{m}{2m-1}s and \AME{2m,d} states \cite{Helwig2012, Helwig2013}. The dimension $d$ of the systems in the AME state translate to a $d$-dimensional secret and $d$-dimensional share sizes for each player in the QSS scheme. Here, we want to show how this construction of threshold QSS schemes from AME states work in the presented graph state formalism.

Given an \AME{2m,p} graph state $\ket{G}$, the role of the \emph{dealer} $D$ is assigned to one of the $2m$ parties. The dealer possesses an additional state, the secret $\ket{s} = \sum \alpha_i \ket{i}$, and his job is to encode this secret onto the qudits shared by the other $2m-1$ players. He does that by performing a generalized Bell measurement, which is a projective measurement onto the basis
\begin{equation}
 \label{eq:Bell}
 \ket{\Psi_{gh}} = \frac{1}{\sqrt{p}}\sum_j e^{2\pi ijg/p} \ket{j}\ket{j+h},
\end{equation}
on the secret and his qudit of the graph state. This results in the encoded state
\begin{align}
 \ket{\Phi_S} = \sum_{i=0}^{p-1} \beta_i \ket{G^{\backslash D}_{iA_D\backslash \{D\}}},
\end{align}
where $\beta_i = \braket{i|U^\dagger_{gh}|s}$, with
\begin{align}
	\label{eq:BellU}
	U_{gh} &= \sum_j e^{2\pi ijg/p} \ket{j}\bra{j+h},
\end{align}
depends on the outcome of the Bell measurement $(g,h)$. This outcome has to be broadcasted to the remaining $2m-1$ players $P$. To see that the resulting state is a \QSS{m}{2m-1}, we have to confirm that any subset $B$ of $m$ players can recover the secret. Tracing out the other $m-1$ parties $K=P\backslash B=\{k_1,\ldots,k_{m-1}\}$ gives
\begin{align}
 \rho &= \Tr_K \ket{\Phi_S}\bra{\Phi_S}\\
  &= \sum_{i,j} \sum_{\vv a \in \ZZ_p^{m-1}}
     \beta_i \beta^\ast_j \braketind{D,K}{i,a_1,\ldots, a_{m-1}|G}{}
     \braketind{}{G|j,a_1,\ldots, a_{m-1}}{D,K}\\
  &= \sum_{i,j} \sum_{\vv a \in \ZZ_p^{m-1}}
     \beta_i \beta^\ast_j \ket{G^{\backslash \{D,K\}}_{iA_D + \sum_l a_l A_{k_l}\backslash \{D,K\}}}
     \bra{G^{\backslash \{D,K\}}_{j A_D + \sum_l a_l A_{k_l}\backslash \{D,K\}}}.
\end{align}
Since the vectors $\{A_D\backslash \{D,K\}, A_{k_1}\backslash \{D,K\}, \ldots, A_{k_{m-1}}\backslash \{D,K\}\}$ are linearly independent,
\begin{equation}
 V: \ket{G^{\backslash \{D,K\}}_{iA_D + \sum_l a_l A_{k_l}\backslash \{D,K\}}}
 \rightarrow \ket{i,a_1,\ldots, a_{m-1}}
\end{equation}
is a unitary operation on the qudits shared by the players in $B$. Applying it to $\rho$ gives
\begin{align}
 V\rho V^\dagger 
  &= \ket{s'}\bra{s'} \otimes \sum_{\vv a} \ket{a_1,\ldots, a_{m-1}}\bra{a_1,\ldots, a_{m-1}},
\end{align}
with $\ket{s'} = \sum_i \beta_i \ket{i}$. Thus after applying $U_{gh}$ to the first qudit, the secret is restored. 

That any set with less than $m$ players is forbidden follows directly from the no-cloning theorem. Thus we can construct a \QSS{m}{2m-1} with graph states from any \AME{2m,d} graph state.

%

\subsection{Ramp QSS Schemes}
A generalization of threshold secret sharing schemes are $(m,L,n)$ \emph{ramp secret sharing schemes} \cite{Blakley1984}. In these schemes $n$ players share a state such that any set of $m$ or more players can recover the secret and any set of $m-L$ or less players is a forbidden set, while any set in between is an intermediate set. The special case of $L=1$ is a threshold secret sharing scheme.

It was shown in Ref.~\cite{Helwig2013} that a $(m,L,2m-L)$ ramp QSS scheme can be constructed from an \AME{2m,d} state for all $1\leq L \leq m$. In this scenario each of the $2m-L$ players possesses a system of dimension $d$, while the dimension of the secret is $d^L$. 
This is achieved by assigning the role of the dealer to more than one party in the previously presented threshold QSS scheme. This method also works in the graph state formalism. Note that in this scenario, contrary to the the threshold scheme presented earlier, the secret dimension can be larger than the system of each player. This is achieved by having a ``weaker'' security structure with intermediate sets.

Consider an \AME{2m,p} graph state. We assign $L$ dealers $D=\{d_1, \ldots, d_L\}$. Each of them performs a bell measurement on their qudit of the graph state an a secret $\ket{s_m} = \sum_i \alpha_{m,i} \ket{i}$. Without loss of generality we assume that the measurement result is $(0,0)$, different measurement outcomes could be corrected in the end in the same way as for the threshold QSS scheme. After the Bell measurements, the remaining $2m-L$ players $P$ share the state
\begin{equation}
 \ket{\Phi_S} = \sum_{i_1, \ldots, i_L} \alpha_{1,i_1}\cdots \alpha_{L,i_L}
    \ket{G^{\backslash D}_{\sum_m i_m A_{d_m}\backslash D}}.
\end{equation}
Now any subset $B\subset P$ of $m$ players should be able to recover the secret. Tracing out $K=P\backslash B=\{k_1,\ldots,k_{m-L}\}$ gives
\begin{align}
 \rho =& \Tr_K \ket{\Phi_S}\bra{\Phi_S}\\
  =& \sum_{\substack{i_1,\ldots, i_L\\ j_1,\ldots, j_L}}
     \sum_{\vv a \in \ZZ_p^{m-L}}
     \alpha_{1,i_1}\cdots \alpha_{L,i_L}
     \alpha^\ast_{1,j_1}\cdots \alpha^\ast_{L,j_L}\\
     &\ket{G^{\backslash \{D,K\}}_{\sum_m i_m A_{d_m} + \sum_l a_l A_{k_l}\backslash \{D,K\}}}
     \bra{G^{\backslash \{D,K\}}_{\sum_m j_m A_{d_m} + \sum_l a_l A_{k_l}\backslash \{D,K\}}}.
\end{align}
And applying $V$ recovers the secrets:
\begin{equation}
 V\rho V^\dagger = \ket{s_1}\bra{s_1} \otimes \cdots \otimes \ket{s_L}\bra{s_L}
    \otimes \sum_{\vv a} \ket{a_1,\ldots, a_{m-L}}\bra{a_1,\ldots, a_{m-L}}
\end{equation}
That any subset of $m-L$ players or less cannot gain any information about the secrets follows again from the no-cloning theorem. For a discussion while sets of players with more than $m-L$ but less than $m$ players are indeed intermediate sets, which means they cannot recover the full secrets, but gain some information, see Ref.~\cite{Helwig2013}.

\section{Conclusion and Open Questions}
\label{section:conclusion}
In this paper, we have shown how the graph state formalism can be used to describe absolutely maximally entangled states. Due to the high degree of multipartite entanglement in graph states, they provide an optimal framework for the investigation of AME states.
Furthermore, they intrinsically provide a quantum circuit to generate the state, which should be very useful for actually implementing AME states in the future, once it is possible to experimentally design CNOT gates for qudits.
Two different methods to check bipartite entanglement in graph states have been presented. One uses a graphical illustration of the existing entanglement in the graph, while the other one provides a very efficient method to check if a given graph state is absolutely maximally entangled.

With the efficient method, we are able to numerically check the entanglement of millions of graph states per minute, which we were able to use to find a previously unknown AME state for seven qutrits. Unfortunately, with increasing system dimensions and number of parties, the number of possible graph states grows exponentially, which makes an exhaustive search already infeasible for eight qutrits. Hence for future investigations, a goal would be to combine both methods. Using insight gained from the graphical representation might help us cut down on the number of graph states that are candidates for AME states.

In addition to the seven qutrit AME graph state, we were able to construct an AME graph state for all previously known AME states, in particular for each number of parties, an AME graph state can be constructed from classical MDS codes. Thus the question arises if we can always find an \AME{n,d} graph state if an \AME{n,d} state exists. So far, we were not able to either proof that or construct a counterexample.

Finally, we showed how AME graph states can be used for quantum secret sharing within the graph states formalism. QSS with graph states has already been introduced before \cite{Markham2008, Keet2010}, but only two examples for threshold QSS schemes for 4 and 6 qudits, corresponding to the graph states in Figures~\ref{fig:AME43state} and \ref{fig:AME62wheel}, respectively, were given. However, it remained an open question, with other graph states are suitable for threshold QSS schemes.
Here we showed that all AME graph states shared between an even number of parties can be used to derive threshold QSS schemes, as well as ramp QSS schemes, which have not been covered before in the graph state formalism.

\section*{Acknowledgements}
I acknowledge financial support by NSERC and the CRC program. I also want to thank Hoi-Kwong Lo and David Gosset for helpful discussions and comments.

\bibliographystyle{ieeetr}
\bibliography{entanglement}

\end{document}